\documentstyle[11pt,paspconf,epsf]{article}

\begin{document}  

\title{The Lyman Alpha Forest Within The Cosmic Web}

\author{J. Richard Bond and James W. Wadsley}
\affil{Canadian Institute for Theoretical Astrophysics and 
Department of Astronomy, University of Toronto, 
McLennan Physical Laboratories, Toronto,
ONT M5S 1A1, Canada}

\begin{abstract}
Observations indicate galaxies are distributed in a filament-dominated
web-like structure; classic examples are the Coma and Perseus-Pisces
superclusters. Numerical experiments at high and low redshift of
viable structure formation theories also show filament-dominance; in
particular, the gasdynamical simulations of Ly$\alpha$ clouds at
redshifts $\sim 2-6$ that we concentrate on here. We understand why
this is so in terms of rare events (peak patches) in the medium and
the web pattern of filaments that bridge the gaps between the peaks
along directions defined by their (oriented) tidal fields. We present
an overview of these ideas and their practical application in crafting
high resolution well-designed simulations. We show the utility of this
by taking a highly filamentary subvolume found in a galactic-scale
simulation, compressing its important large scale features onto a
handful of numbers defining galactic-scale peak-patch constraints,
which are then used to construct constrained initial conditions for a
higher resolution simulation appropriate for study of the
Ly$\alpha$ forest.
\end{abstract}

\section{The Peak Patch Picture and the Cosmic Web Theory of Filaments}

We approach the connected ideas we loosely call Cosmic Web theory
(Bond, Kofman and Pogosyan 1996, BKP) and the Peak Patch Picture (Bond
and Myers 1996, BM, refer here for pre-1995 work) through a historical path
that includes an outline of the relevant terminology. In 1965, Lin,
Mestel \& Shu showed that a cold triaxial collapse implied an oblate
``pancake'' would form.  In 1970, Zeldovich developed his famous
approximation and argued that pancakes would be the first structures
that would form in the adiabatic baryon-dominated universes popular at
that time. Generally for a cold medium, there is a full non-linear
map: ${\bf x} ({\bf r} , t) \equiv {\bf r}-{\bf s} ({\bf r} , t)$,
from Lagrangian (initial state) space, ${\bf r}$, to Eulerian (final
state) space, ${\bf x}$. The map becomes multivalued as nonlinearity
develops in the medium. It is conceptually useful to split the
displacement field, ${\bf s} = {\bf s}_b + {\bf s}_f$, into a smooth
quasilinear long wavelength piece ${\bf s}_b$ and a residual highly
nonlinear fluctuating field ${\bf s}_f$. If the {\it rms} density
fluctuations smoothed on scale $R_b$, $\sigma_\rho (R_b)$, are $<
{\cal O}(1/2)$, the ${\bf s}_b$-map is one-to-one (single-stream)
except at the rarest high density spots. In the peak patch approach,
$R_b$ is adaptive, allowing for dynamically hot regions like
protoclusters to have large smoothing and cool regions like voids to
have small smoothing. If $D(t)$ is the linear growth factor, then
${\bf s}_b =D(t) {\bf s}_b ({\bf r} ,0 )$ describes Lagrangian linear
perturbation theory, {\it i.e.} the Zeldovich approximation. The large
scale peculiar velocity is ${\bf V}_{Pb} = -\bar{a}(t) \dot{{\bf
s}}_b({\bf r}, t)$. What is important for us is the strain field (or
deformation tensor): { \small
\begin{eqnarray}
&& e_{b,ij} ({\bf r})\equiv - {1\over 2} \big({\partial s_{bi} \over
\partial r_j}+ {\partial s_{bj} \over \partial r_i}\big) ({\bf r} ) =
-\sum_{A=1}^3 \lambda_{vA} {\hat n}_{vA}^{i} {\hat n}_{vA}^{j},{\rm~where}\nonumber 
\label{eq:eb}\\ && \
\lambda_{v3}={\delta_{Lb}\over 3}(1+3 e_v + p_v),\
\lambda_{v2}={\delta_{Lb}\over 3}(1-2 p_v),\
\lambda_{v1}={\delta_{Lb}\over 3}(1+3 e_v - p_v), \nonumber 
\end{eqnarray}}and $\delta_{Lb}=-e_{b,i}^i$ is the smoothed linear
overdensity, which we often express in terms of the height relative to
the {\it rms} fluctuation level $\sigma_\rho (R_b)$, $\nu_b \equiv
\delta_{Lb}/\sigma_\rho (R_b)$. The deformation eigenvalues are
ordered according to $\lambda_{v3}\ge \lambda_{v2} \ge \lambda_{v1}$
and ${\hat n}_{vA}$ denote the unit vectors of the principal axes. In
that system, $x_A=r_A (1-\lambda_{vA} ({\bf r} , t))$ describes the
local evolution. The Zeldovich-mapped overdensity is $
(1+\delta_Z)({\bf r} ,t )= \vert {(1-D(t)\lambda_{v3}
)(1-D(t)\lambda_{v2} )(1-D(t)\lambda_{v1})}\vert^{-1}$, exploding when
the largest eigenvalue $D(t)\lambda_{v3}$ reaches unity (fold caustic
formation). In a Zeldovich map a pancake developes along the surface
${\hat n}_{v3} \cdot \nabla_{{\bf r}} \lambda_{v3} =0$.

The strain tensor is related to the peculiar linear tidal tensor by
${\partial^2 \Phi_P \over \partial x^i \partial x^j} = - 4\pi G {\bar
\rho}_{nr}\bar{a}^2\, e_{b,ij}$, where $\Phi_P$ = peculiar
gravitational potential, and to the linear shear tensor by ${\dot
e}_{b, ij}$. The anisotropic
part of the shear tensor has two independent parameters, the
ellipticity $e_v$ (always positive) and the prolaticity $p_v$. 

Doroshkevich (1973) and later Doroshkevich \& Shandarin (1978) were
among the first to apply the statistics of Gaussian random fields to
cosmology, in particular of $\lambda_{vA}$, at random points in the
medium. Arnold, Shandarin and Zeldovich (1982) made the important step
of applying the catastrophe theory of caustics to structure
formation. This work suggested the following formation sequence:
pancakes first, followed by filaments and then clusters.  This should
be compared to the BKP Web picture formation sequence: clusters first,
followed by filaments and then walls.  BKP also showed that filaments
are really ribbons, walls are webbing between filaments in cluster
complexes, and that walls are not really classical pancakes.  For the
Universe at $z\sim 3$, massive galaxies play the role of clusters, and
for the Universe at $z\sim 5$ more modest dwarf galaxies take on that
role.

The Web story relies heavily upon the theory of Gaussian random fields
as applied to the rare ``events'' in the medium, {\it e.g.}, high density
peaks. Salient steps in this development begin with Bardeen {\it et
al.} (1986, BBKS), where the statistics of peaks were applied to
clusters and galaxies, {\it e.g., } the calculation of the peak-peak
correlation function, $\xi_{pk,pk}$.  In a series of papers, Bond
(1986-90) and Bond \& Myers (1990-93) developed the theory so that it
could calculate the mass function, $n(M)dM$.  It was also applied to
the study of how shear affects cluster alignments ({\it e.g., } the
Binggeli effect), and to Ly$\alpha$ clouds, `Great Attractors', giant
`cluster-patches', galaxy, group and cluster distributions, dusty PGs,
CMB maps and quasars. This culminated in the BM ``Peak-Patch Picture
of Cosmic Catalogues''.

\begin{figure}
\centerline{
\epsfxsize=2.5in\epsfbox{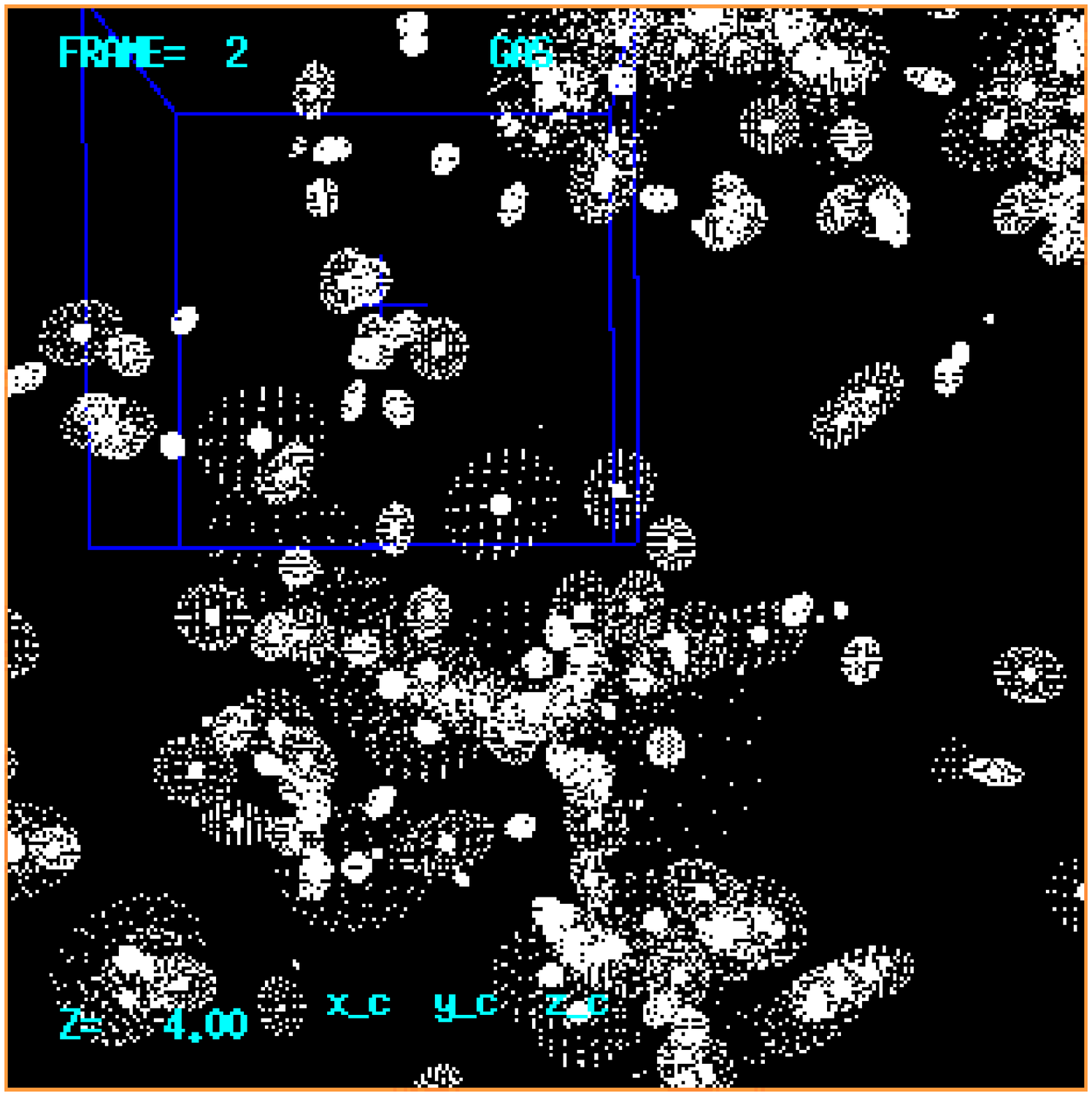}
\hspace{0.0in}
\epsfxsize=2.5in\epsfbox{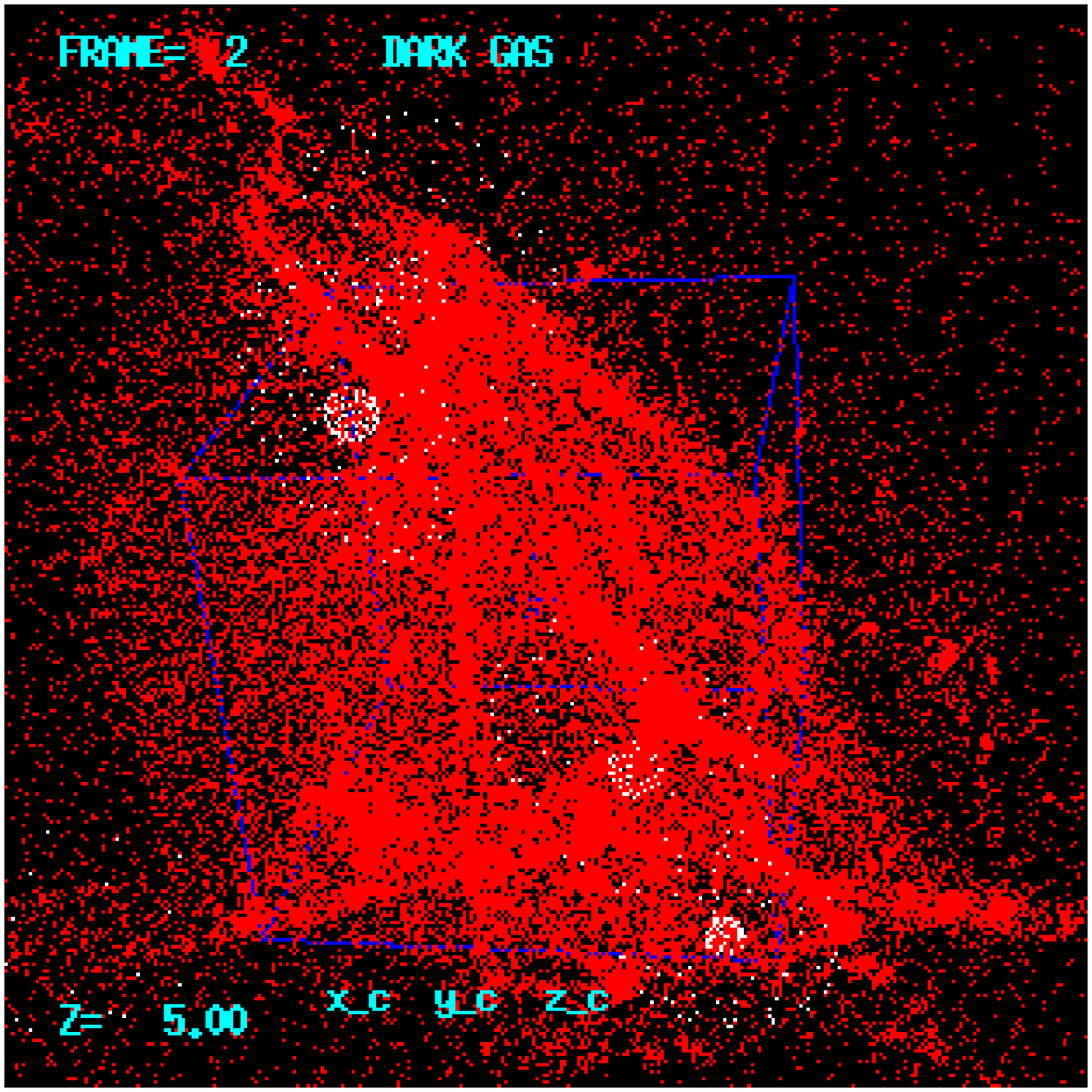}}
\vspace{0.0in}
\centerline{
\epsfxsize=2.5in\epsfbox{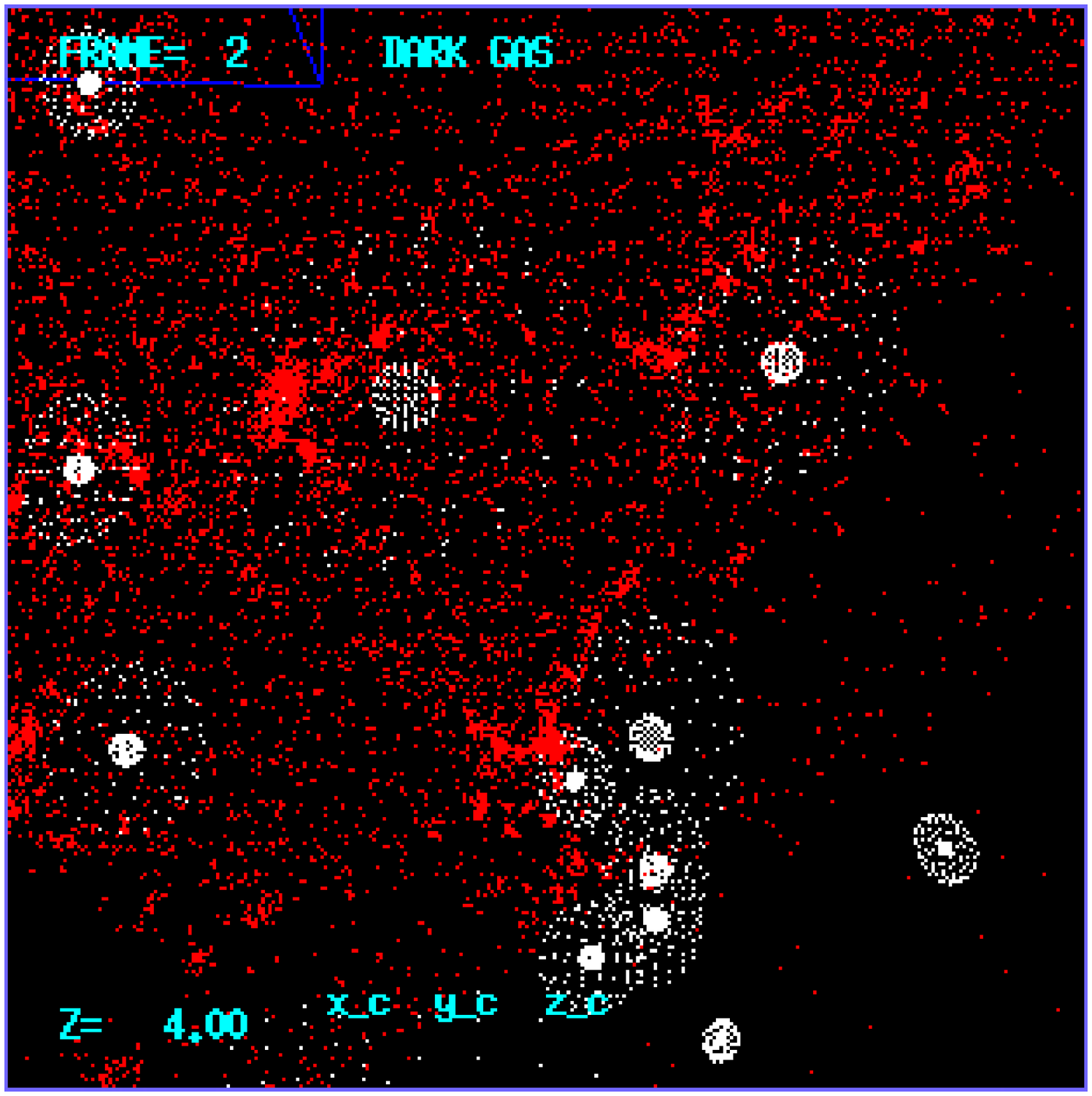}
\hspace{0.0in}
\epsfxsize=2.5in\epsfbox{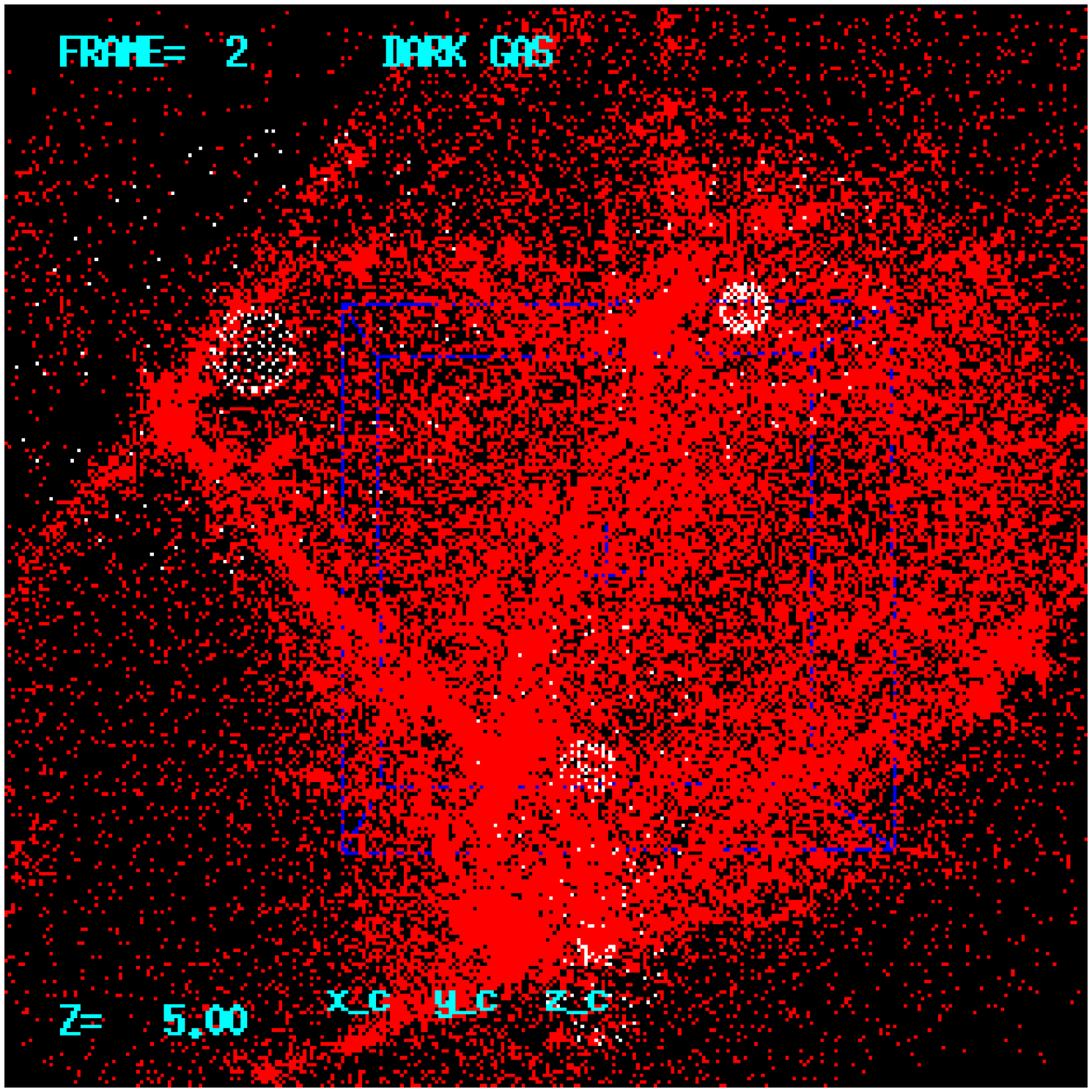}}
\caption{\small These plots show the reconstruction of a galaxy-galaxy
filament present in our 40\,Mpc ``galaxy'' simulation. The cosmology
is a standard initially scale-invariant $\Omega_{nr}$=1 CDM model with
$\Omega_{B}$=0.05, ${\rm h}$=0.5, normalized to $\sigma_8$=0.67.  We
use the cosmological SPH+P$^3$MultiGrid code described in WB. We
identified the peak-patches that should collapse by $z=4$ in the
initial conditions (IC), as shown in the top left panel (which is
10\,$h^{-1}$Mpc across, comoving). For each patch, the outer ellipsoid
represents the alignment of the shear tensor and the inner sphere is
an estimate of the final object size. In the lower left panel, we zoom
in on the central filamentary web structure and overlay dark matter
from a simulation of these IC (panel 2.5\,$h^{-1}$Mpc across). The
five peak patches (at $z=4$ with overdensity 180) and two voids that
define this region were used as constraints for a new higher
resolution IC (12.8\,Mpc), which we also evolved numerically. These
peaks are shown overlaid with dark matter from the new simulation in
the right hand panels. The top right panel is a different orientation
to the others that shows the filament more clearly, also shown in
$n_{HI}$ in Fig.~2. These panels demonstrate that peaks represent an
excellent way to compress the essential information about large scale
filamentary behaviour.  }
\label{dotplot} 
\end{figure}

We briefly describe the BM peak patch method and how it is applied to
initial conditions for simulations; an example is shown in
Fig.~\ref{dotplot}. We identify candidate peak points using a
hierarchy of smoothing operations on the linear density field
$\delta_{L}$. To determine patch size and mass we use an ellipsoid
model for the internal patch dynamics, which are very sensitive to the
external tidal field. A byproduct is the internal (binding) energy of
the patch and the orientation of the principal axes of the tidal
tensor. We apply an exclusion algorithm to prevent peak-patch
overlap. For the external dynamics of the patch, we use a
Zeldovich-map with a locally adaptive filter ($R_{pk}$) to find the
velocity ${\bf V}_{pk}$ (with quadratic corrections sometimes
needed). The peaks are rank-ordered by mass (or internal
energy). Thus, for any given region, we have a list of the most
important peaks.  By using the negative of the density field, we can
also get void-patches.  Some of the virtues of the method are: it
represents a natural generalization of the Press-Schechter method to
include non-local effects;\footnote{The hugely popular,
trivial-to-implement, Press-Schechter (1976) method for determining
$n(M)$ has been the principal competitor to the peaks theory over the
years, but it has no real physical basis and disagrees strongly with
the spatial distribution (Bond {\it et.al.} 1991); thus the amazement,
and delight, in the community that $n(M)$ fits that of $N$-body group
catalogues so well.}  is a natural generalization of BBKS
single-filter peaks theory to allow a mass spectrum and solve the
cloud-in-cloud ({\it i.e.}, peak-in-peak) problem; allows efficient
Monte Carlo constructions of 3D catalogues; gives very good agreement
with $N$-body groups; has an accurate analytic theory with which to
estimate peak properties, ({\it e.g., } mass and binding energy from
mean-profiles, using $\delta_{L,crit}(e_v)$, $\langle{e_v \vert
\nu_{pk}\rangle}$); and handles merging, with high redshift peaks
being absorbed into low redshift ones.

BKP concentrated on the impact the peak-patches would have on their
environment and how this can be used to understand the web. They
showed that the final-state filament-dominated web is present in the
initial conditions in the $\delta_{Lb}$ pattern, a pattern
largely determined by the position and primordial tidal fields of
rare events. BKP also showed how 2-point rare-event constraints define
filament sizes (see their Fig. 2). The strongest filaments are
between close peaks whose tidal tensors are nearly aligned. Strong
filaments extend only over a few Lagrangian radii of the peaks they
connect. They are so visually impressive in Eulerian space because the
peaks have collapsed by about 5 in radius, leaving the long bridge
between them, whose transverse dimensions have also decreased. This is
illustrated by the lower right panel of Fig.~\ref{dotplot} in which
the aligned galaxy peaks are connected by strong filaments. Strong
vertical filaments are a product of the vertical alignment of the
peaks' tidal tensors, which simultaneously acts to prevent a strong
horizontal filament between the top two peaks. The reason for this
phenomenon is that the high degree of constructive interference of the
density waves required to make the rare peak-patches, and to
preferentially orient them along the 1-axis, leads to a slower
decoherence along the 1-axis than along the others, and thus a higher
density. 3-point and higher rare-event constraints of nearby peaks
determine the nature of the webbing between the filaments, also
evident in Fig.~\ref{dotplot}.

\section{Crafting High Resolution $N$-body/Hydrodynamical Simulations}

Although it is usual to evolve ambient ``random'' patches of the
Universe in cosmology, there are obvious advantages in spending one's
computational effort on the regions of most interest. Single peak
constraints are very useful if cluster or galaxy formation is the
focus, while multiple peak constraints are more useful if
superclusters, or cluster substructure, or filaments and walls are the
focus. We have seen that the essential features at a given epoch of
the filamentary structure and the wall-like webbing between the
filaments is largely defined by the dominant collapsed structures, and
the peak patches that gave rise to them. A general method for building
peak environments is suggested: construct random field initial
conditions that require the field to have prescribed values of the
peak shear (smoothed over the peak size at the peak position), for a
subset from the size-ordered list of peaks that will have a strong
impact on the patch to be simulated. Only a handful $N_{pk}$ of peaks
and/or voids is usually needed to determine the large scale features,
effectively compressing the information needed to $(3+1+6)N_{pk}$
numbers (${\bf r}_{pk}, M_{pk}, e_{pk,ij}$); peak velocities and the
peak constraint are also usually added but these are not as important.

To illustrate how this works, we created a random (unconstrained)
initial state for a CDM model in a 40\,Mpc box, our ``galaxy''
simulation. We found peak patches according to the method described in
BM and focussed on a specific subregion exhibiting a strong filament,
choosing the peaks and voids that were expected to exert strong tidal
influences within and upon the patch, as described in
Fig.~\ref{dotplot}. The region chosen was just above the large central
cluster of peaks in the top left panel of Fig.~\ref{dotplot}. We
constructed a higher resolution set of ``Ly$\alpha$ cloud'' initial
conditions for this patch, which the ``galaxy'' initial condition
simulation could not resolve well enough to address the low column
depth Ly$\alpha$ forest of interest to us. 

\begin{figure}
\epsfxsize=\hsize
\centerline{
\epsfxsize=5.0in\epsfbox{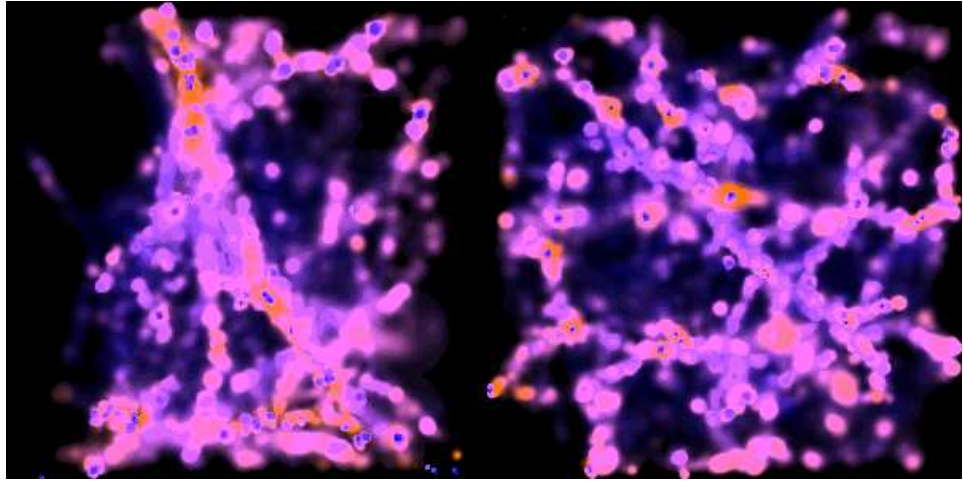}}
\caption{ \small Left: A pseudo ray-traced rendering of the HI density
for a region at $z$=4 constrained by the complex pattern of
galaxy-scale peak and void shear-fields described in the text,
encoding the main large scale structure features, and which ensures
strong filamentary webbing arises in the patch. The colour scale codes
the temperature variation, with the darker shades in the centre of
objects representing high column cold gas. The region is
$2.5\,h^{-1}Mpc$ across, comoving, and corresponds to the top right
panel of fig.~1. Right: $n_{HI}$ for an ambient patch of the Universe
at $z$=3 with control parameter $\nu_b$=0, simulation as described in
WB. Note the filament dominance in both cases.}
\label{ray} 
\end{figure}

By compressing the initial data in our target region to just the
positions and shears of a few large peak patches, then forming a
constrained realization and applying different random waves
(optimally-sampled for the smaller region) than the original $40 Mpc$
initial condition used, we know we will get high frequency structure
wrong. But clearly the large scale features are the same. This is in
spite of the tremendously complex filamentary structure just below our
chosen sub-region. The peaks we chose were on the basis of rareness
(size) and proximity to the patch (using an algorithm roughly based on
correlation function falloff from each peak).  The five peak-patches
used for this companion Ly$\alpha$ simulation had the following masses
(in units of $10^{11}{\rm\,M_\odot}$) and halo velocity dispersion (in
units of $km\,s^{-1}$) as determined from the binding energy: $3.6,
77$; $3.5, 80$; $1.4, 57$; $0.85,48$; $0.51, 40$.  These accord well
with what our group finder finds in the simulation at this
redshift. The two void-patches used had Lagrangian masses of $2.4$ and
$0.72$, and were outside the high resolution interior.  The
approximate alignments of the shear tensors for the peak patches
inside ensure that a strong filament exists. Fig.~\ref{ray} shows how
the filament looks in HI column density.

\begin{figure}
\vspace{-.5in}
\centerline{\epsfxsize=5.in\epsfbox{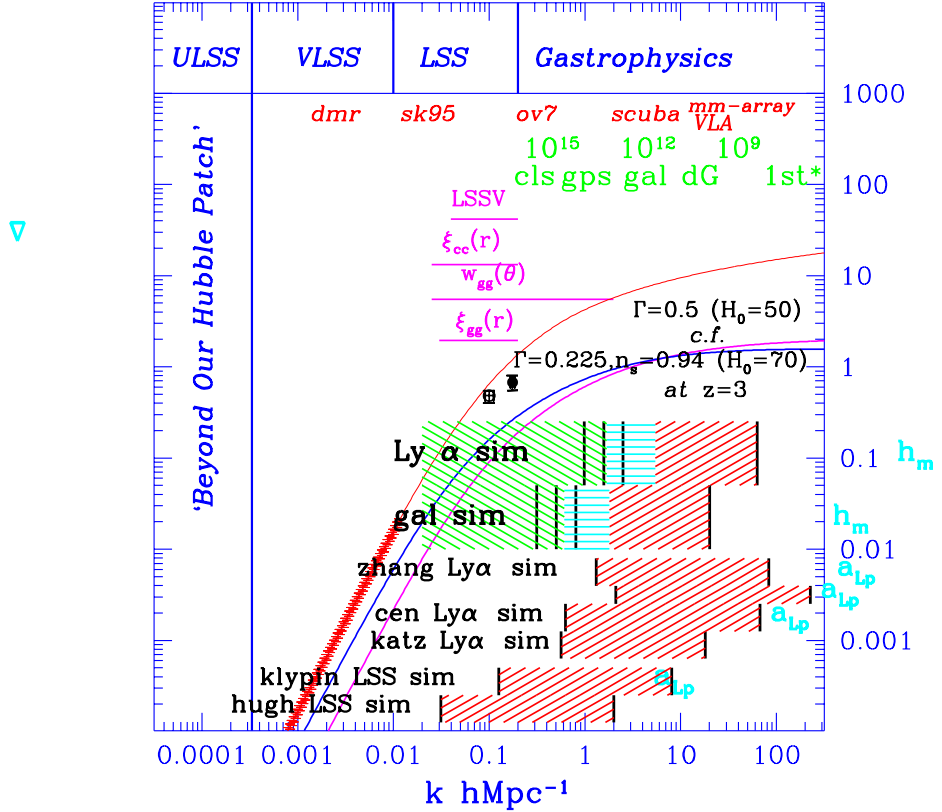}}
\vspace{-.5in}
\caption{\small The two (linear) power spectra shown (scaled to
redshift 3) were used in the simulations here and in WB. The upper
curve at high $k$ is the standard untilted CDM model, but normalized
to cluster abundances, $\sigma_8=0.67$. The other has the same
cosmological age (13 Gyr) and $\Omega_B{\rm h}^2$ (0.0125) but
$H_0=70$ with $\Omega_\Lambda = 0.67$, slightly tilted to be
COBE-normalized ($n_s=0.94$).  (Also shown is a COBE-normalized CDM
model, which misses the solid data point in the cluster-band
(constraint from $dn_{cl}/dT_X$).) The bands in comoving wavenumber
probed by various simulations are contrasted. Periodic simulations may
use the entire volume, but the $k$-space restriction to lie between
the fundamental mode (low-$k$ boundary line) and the Nyquist
wavenumber (high-$k$ boundary line) in the IC can severely curtail the
rare events in the medium that observations especially probe, and
prevents tidal distortions of the simulation volume. We use 3
$k$-space sampling procedures (FFT and two direct FTs) with the
boundaries defined by which has the smallest volume per $k$-mode. Even
though a $256^3$ Fourier transform was used, notice how early the
direct sampling takes over (with only 10000 modes). Using an FFT with
the very flat spectra in the dwarf galaxy (dG) band can give
misleading results. The 3 low-$k$ lines shown for our Ly$\alpha$ and
galaxy simulations correspond to the high, medium and low resolution
fundamental modes. We actually include modes in the entire hatched
region, with the tidal fields associated with the longer waves
included by a self-consistent uniform tide on the LR simulation
volume. $h_m$ denotes our best resolution. $a_{Lp}$ denotes the
physical (best) lattice spacing for the grid-based Eulerian hydro
codes of Cen and Zhang {\it et.al.} ($z=3$). $k$-space domains for two
large scale structure ($z=0$) simulations are also shown, a Klypin
$256^3$ $PM$ calculation and a Couchman ({\it hugh}) $128^3$ $P^3M$
simulation.  }
\label{probes} 
\end{figure}

To do such complex regions, the standard periodic box approach to
cosmological simulations is obviously not appropriate.  Advantages of
focussed non-periodic simulations include: (1) good mass resolution,
allowing us to concentrate on the scale needed to adequately treat the
objects that form ({\it e.g.} dwarf galaxies,
$a_L=a_{Lattice,High\,Res}=100$\,kpc), and corresponding high
numerical resolution ($h_{sph,min}$, $h_{grav}$ $\sim 1$ {\rm\,kpc}),
can be achieved; (2) good $k$-space sampling is also possible. The
competing demands of $k$-space sampling and resolution are further
described in Fig.~\ref{probes}. Our method allows high resolution
without compromising our long wave coverage by going beyond grid based
FFTs, with a FastFT for high $k$ (which kicks out well before the
fundamental mode is reached) that is superseded by two direct FTs,
power-law then log $k$ sampling, with transitions among them 
determined by minimizing the volume per mode in
$k$-space. Well-sampled $k$-space is especially important for
Ly$\alpha$ cloud and galaxy formation as opposed to cluster
formation because the density power spectrum for viable hierarchical
theories has nearly equal power per decade (approaching flatness in
Fig.~\ref{probes}): if just the FFT is used, as is often the case in
cosmology even for non-periodic calculations, because there is only
one fundamental $k$-mode along each box axis, the large scale
structure in the simulation will be poorly modelled, and this can also
have a deleterious effect on small scale structure.

There is no point adding long waves without maintaining accurate large
scale tides and shearing fields during the calculation. We achieve
this with a high resolution region of interest (grid spacing $a_L$,
$50^3$ sphere) that sits within a medium resolution region ($2\,a_L$,
$40^3$), in turn within a low resolution region ($4\,a_L$,
$32^3$). The influence of ultra long waves is included by measuring
the mean external tide acting on the low resolution region in the
initial conditions, adopting simple models for the ultra-long wave
dynamics based on that measurement ({\it e.g.} linear, Zeldovich, or
homogeneous ellipsoid, as in BM) and applying it as an ``external
force'' throughout the simulation. For this simulation, linear ultra
long wave dynamics were adequate.

In WB, instead of complex multipeak constraints for individual
regions, we use {\it importance sampling} of {\bf shearing patches}
(patches with the smoothed shear tensor $e_{b,ij}(0)$ prescribed at
the centre) to maximize the statistical information we can get from a
crafted set of relatively modest constrained-field SPH calculations,
defined by a set of control parameters, here the central $\nu_b$,
$e_v$, $p_v$ smoothed over a galactic-scale $R_b$. This allows us to
sample rare peak and void patches, difficult to sample even in large
box simulations, especially if FFTs are used. These are in addition to
patches with more typical {\it rms} density contrasts.  We then
combine the results to get the frequency distribution of, say,
$N_{HI}$ for a random patch of the Universe using Bayes theorem, which
decomposes it into the frequency distribution for $N_{HI}$ for our
constrained patches given the control parameters, measured from the
simulations, and the known probability distribution of the control
parameters: schematically,

\noindent 
{\small $P({{\tt random-patch}} ) = \int P( {\tt constr-patch} \vert
{\tt control} ) \ P({\it \tt control} ) \ d {\it \tt control-param}$}.

\noindent
Shearing patches with low $\vert \nu_b \vert $ $
{{\raisebox{-.7ex}{$<$}}\atop{\raisebox{+.2ex}{$\sim$}}} 2$ often have
relatively large shear ellipticity, $\langle e_v \, \vert \nu_b
\rangle \approx 0.54\vert \nu_b \vert^{-1}$, which can give strongly
asymmetric collapses for $\nu_b > 0$, amplifying the smaller scale
filamentary webbing by concentrating it in larger scale filaments or
walls (which depends upon $p_v$). Fig.~\ref{ray} shows that a
$\nu_b=0$ shearing patch has its $n_{HI}$ filaments more spread out
than the multipeak case we have described here, which is reminiscent
of, but even more concentrated than, a $\nu_b=1.4$ calculation shown
in WB.

Simulating a large number of controlled patches in parallel is a form
of adaptive refinement, of which there is much discussion in these
proceedings. In refined regions, the initial conditions are almost
never modified with higher frequency waves, so the Lagrangian ({\it
i.e.} mass) resolution remains fixed even though the Eulerian
resolution may be superb. (This is especially vexing for voids.) When
we refine a region by creating a high resolution realization with the
information contained in peak patches, we {\it optimally} resample
$k$-space to generate a new set of high frequency waves. It is clear
that the cosmological codes of the future will have to simultaneously
adapt in Eulerian and $k$ space, and the techniques explored here
offer a promising path towards this goal.  

\acknowledgments

Support from the Canadian Institute for Advanced Research and NSERC is
gratefully acknowledged. We thank Lev Kofman, Steve Myers and Dmitry
Pogosyan for much fun peak-patch/web interaction.

\end{document}